# Research Challenges in Wireless Sensor Network: A State of the Play

Sukhchandan Randhawa[#1]

[#1]Computer Science and Engineering Department, Thapar University, Patiala, India
[1]sukhchandan@thapar.edu

**Abstract**
A wireless sensor network (WSN) has important applications such as remote environmental monitoring and target tracking. This has been enabled by the availability, particularly in recent years, of sensors that are smaller, cheaper, and intelligent. These sensors are equipped with wireless interfaces with which they can communicate with one another to form a network. The design of a WSN depends significantly on the application, and it must consider factors such as the environment, the application's design objectives, cost, hardware, and system constraints. The goal of survey is to present a comprehensive review of the recent literature in wireless sensor network. This paper reviews the major development and new research challenges in this area.

**Keywords**
Wireless sensor network, Protocols, Sensor network services, Sensor network deployment, Survey

## 1. Introduction

Wireless sensor networks (WSNs) have gained worldwide attention in recent years, particularly with the proliferation in Micro-Electro-Mechanical Systems (MEMS) technology which has facilitated the development of smart sensors. These sensors are small, with limited processing and computing resources, and they are inexpensive compared to traditional sensors [1]. These sensor nodes can sense, measure, and gather information from the environment and, based on some local decision process, they can transmit the sensed data to the user. The availability of low-cost hardware such as CMOS cameras and microphones has fostered the development of Wireless Multimedia Sensor Networks (WMSNs), i.e., networks of wirelessly interconnected devices that are able to ubiquitously retrieve multimedia content such as video and audio streams, still images, and scalar sensor data from the environment [2].

Wireless multimedia sensor networks will not only enhance existing sensor network applications such as tracking, home automation, and environmental monitoring, but they will also enable several new applications such as:

- Multimedia surveillance sensor networks. Wireless video sensor networks will be composed of interconnected, battery-powered miniature video cameras, each packaged with a low-power wireless transceiver that is capable of processing, sending, and receiving data. Video and audio sensors will be used to enhance and complement existing surveillance systems against crime and terrorist attacks. Large-scale networks of video sensors can extend the ability of law enforcement agencies to monitor areas, public events, private properties and borders [3].

- Storage of potentially relevant activities. Multimedia sensors could infer and record potentially relevant activities (thefts, car accidents, traffic violations), and make video/audio streams or reports available for future query [4].

- Traffic avoidance, enforcement and control systems. It will be possible to monitor car traffic in big cities or highways and deploy services that offer traffic routing advice to avoid congestion. In addition, smart parking advice systems based on WMSNs will allow monitoring available parking spaces and provide drivers with automated parking advice, thus improving mobility in urban areas. Moreover, multimedia sensors may monitor the flow of vehicular traffic on highways and retrieve aggregate information such as average speed and number of cars. Sensors could also detect violations and transmit video streams to law enforcement agencies to identify the violator, or buffer images and streams in case of accidents for subsequent accident scene analysis [5].

- Advanced health care delivery. Telemedicine sensor networks can be integrated with 3G multimedia networks to provide ubiquitous health care services. Patients will carry medical sensors to monitor parameters such as body temperature, blood pressure, pulse oximetry, ECG, breathing activity. Furthermore, remote medical centers will perform advanced remote monitoring of their patients via video and audio sensors, location sensors, motion or activity sensors, which can also be embedded in wrist devices [6].

- Automated assistance for the elderly and family monitors. Multimedia sensor networks can be used to monitor and study the behavior of elderly people as a means to identify the causes of illnesses that affect them such as dementia. Networks of wearable or video and audio sensors can infer emergency situations and immediately connect elderly patients with remote assistance services or with relatives [7].

- Environmental monitoring. Several projects on habitat monitoring that use acoustic and video feeds are being envisaged, in which information has to be conveyed in a time-critical fashion. For example, arrays of video sensors are already used by oceanographers to determine the evolution of sandbars via image processing techniques.

- Person locator services. Multimedia content such as video streams and still images, along with advanced signal processing techniques, can be used to locate missing persons, or identify criminals or terrorists [8].

- Industrial process control. Multimedia content such as imaging, temperature, or pressure amongst others, may be used for time-critical industrial process control. Machine vision is the application of computer vision techniques to industry and manufacturing, where information can be extracted and analyzed by WMSNs to support a manufacturing process such as those used in semiconductor chips, automobiles, food or pharmaceutical products. For example, in quality control of manufacturing processes, details or final products are automatically inspected to find defects [9]. In addition, machine vision systems can detect the position and orientation of parts of the product to be picked up by a robotic arm. The integration of machine vision systems with WMSNs can simplify and add flexibility to systems for visual inspections and automated actions that require high-speed, high-magnification, and continuous operation [10].

## 2. Research Methodology

### 2.1 Research Questions

Novel network protocols that account for the key realities in wireless communication are required. New research is needed to:

1. Measure and assess how the theoretical properties of wireless communication are exhibited in today's and tomorrow's sensing and communication devices,
2. Establish better models of communication realities to feed back into improved simulation tools,
3. Invent new network protocols that account for the communication realities of real world environments,

4. Test the individual solutions on real platforms in real world settings, and Synthesize novel solutions into a complete system-wide protocol stack for a real application.

**2.2 Source of Information**

- ACM Digital Library (<www.acm.org/dl>)
- IEEE eXplore (<www.ieeexplore.ieee.org>)
- ScienceDirect (<www.sciencedirect.com>)
- Google Scholar (<www.scholar.google.co.in>)

**2.3 Research Keywords**

Table 1: Search Keywords and Synonyms

| **Keywords** | **Synonyms** |
|---|---|
| WSN | Wireless sensor network |
| WSN Protocols | Wireless sensor network protocols |
| SNS | Sensor network services |
| SND | Sensor network deployment |
| Survey | Review and Issues |

**2.4 Study Selection**

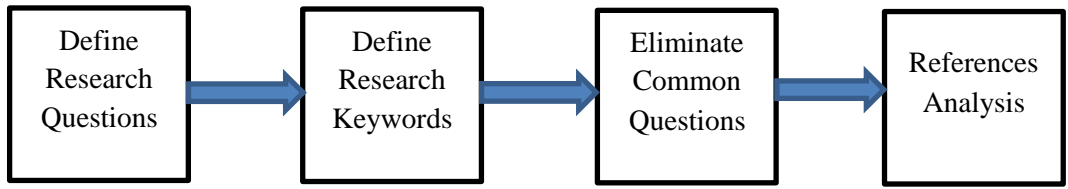

Figure 1: Research Procedure

**3. Analysis**

Few analytical results exist for WSN. Since WSN are in the early stage of development it is not surprising that few analytical results exist. Researchers are busy inventing new protocols and new applications for WSN. The solutions are built, tested and evaluated either by simulation or testbeds; sometimes an actual system has been deployed. Empirical evidence is beginning to accumulate. However, a more scientific approach is required where a system can be designed and analyzed before it is deployed. The analysis needs to provide confidence that the system will meet its requirements and to indicate the efficiency and performance of the system. Consider the following interesting analysis questions.

1. What density of nodes is required to meet the lifetime requirements of the system?
2. What sensing and communication ranges are needed to detect, classify and report a target to a base station by a deadline?
3. What sensing range and what nodes need to be awake in order to guarantee a certain degree of sensing coverage for a system?
4. Given n streams of periodic sensing traffic characterized by a start time, period, message size, deadline, source location and destination location for a given WSN will all the traffic meet their deadlines?

To answer this last question, the interference patterns of wireless communication must be taken into account. Once analysis techniques and solutions are developed for these types of questions, they must also be validated with real systems.

**4. Threats to Validity**

The research papers were obtained by keyword searching and reference analysis. Exclusions were made by reading the title, abstract and conclusions. However, there is a possibility that there exist papers that were missed due to the above searching and exclusion method.

**5. Conclusion and Future Scope**

In this brief note six key research areas were highlighted. However, many other research areas are very important including: localization, topology control, dependability, self-calibration, self-healing, data aggregation, group management, clock synchronization, query processing, sensor processing and fusion under limited capacities, and testing and debugging. WSN are a fascinating area with great potential. The impact of this area on the world can rival the impact that the Internet has had. Exciting and difficult research challenges lie ahead before this becomes reality.